\documentclass[10pt,noshowpacs,twocolumn,superscriptaddress,aps,pra]{revtex4-1}
\usepackage{amsmath}
\usepackage{amssymb}
\usepackage{graphicx}
\DeclareGraphicsExtensions{.pdf,.eps,.png,.jpg,.mps}

\begin{document}
\title{Integrated turnkey soliton microcombs operated at CMOS frequencies}

\author{
Boqiang Shen$^{1,\ast}$,
Lin Chang$^{2,\ast,\dagger}$,
Junqiu Liu$^{3,\ast}$,
Heming Wang$^{1,\ast}$,
Qi-Fan Yang$^{1,\ast}$,\\
Chao Xiang$^2$,
Rui Ning Wang$^3$,
Jijun He$^3$,
Tianyi Liu$^3$,
Weiqiang Xie$^2$,
Joel Guo$^2$,
Dave Kinghorn$^2$,\\
Lue Wu$^1$,
Qing-Xin Ji$^{1,4}$,
Tobias J. Kippenberg$^{3,\dagger}$,
Kerry Vahala$^{1,\dagger}$
and John E. Bowers$^2$\\
$^1$T. J. Watson Laboratory of Applied Physics, California Institute of Technology, Pasadena, CA 91125, USA\\
$^2$ECE Department, University of California Santa Barbara, Santa Barbara, CA 93106, USA\\
$^3$Institute of Physics, Swiss Federal Institute of Technology Lausanne (EPFL), CH-1015 Lausanne, Switzerland\\
$^4$School of Physics, Peking University, Beijing 100871, China\\
$^*$These authors contributed equally to this work.\\
$^\dagger$Corresponding authors: linchang@ucsb.edu, tobias.kippenberg@epfl.ch, vahala@caltech.edu\\
}

\begin{abstract}
While soliton microcombs offer the potential for integration of powerful frequency metrology and precision spectroscopy systems, their operation requires complex startup and feedback protocols that necessitate difficult-to-integrate optical and electrical components. Moreover, CMOS-rate microcombs, required in nearly all comb systems, have resisted integration because of their power requirements. Here, a regime for turnkey operation of soliton microcombs co-integrated with a pump laser is demonstrated and theoretically explained. Significantly, a new operating point is shown to appear from which solitons are generated through binary turn-on and turn-off of the pump laser, thereby eliminating all photonic/electronic control circuitry. These features are combined with high-Q $\mathrm{Si_3N_4}$ resonators to fully integrate into a butterfly package microcombs with CMOS frequencies as low as 15 GHz, offering compelling advantages for high-volume production.
\end{abstract}

\maketitle


\noindent Optical frequency combs have found a remarkably wide range of applications in science and technology \cite{diddams2010evolving}.  And a recent development that portends a revolution in miniature and integrated comb systems is dissipative Kerr soliton formation in coherently pumped high-quality-factor (high-Q) optical microresonators \cite{herr2014temporal,xue2015mode,brasch2016photonic,yi2015soliton,obrzud2017temporal,gong2018high,he2019self}. To date, these soliton microcombs \cite{Kippenberg2018} have been applied to spectroscopy \cite{suh2016microresonator,dutt2018chip,yang2019vernier}, the search for exoplanets \cite{suh2019searching,obrzud2019microphotonic}, optical frequency synthesis \cite{spencer2018optical}, time keeping \cite{newman2019architecture} and other areas. Also, the recent integration of microresonators with lasers has revealed the viability of fully chip-based soliton microcombs \cite{stern2018battery,raja2019electrically}. However, despite their enormous promise, microcomb integration has faced two considerable obstacles.  

First, complex tuning schemes and feedback loops are required for generation and stabilization of solitons \cite{herr2014temporal,yi2016active,joshi2016thermally}. These not only introduce redundant power-hungry electronic components \cite{stern2018battery,raja2019electrically}, but also require optical isolation, a function that has so far been challenging to integrate at acceptable performance levels. Indeed, the omission of optical isolation between a high-Q resonator and a laser has been a subject of study for some time. And semiconductor laser locking to the resonator as well as line narrowing have been shown to result from backscattering of the intracavity optical field\cite{liang2010whispering}. These attributes as well as mode selection when using a broadband pump have also been profitably applied to operate microcomb systems without isolation \cite{liang2015high,pavlov2018narrow,stern2018battery,raja2019electrically}. However, prior studies of feedback effects have considered the resonator to be linear. Here, we show that the nonlinear dynamic of the unisolated laser-microcomb system creates a new operating point from which the pump laser is simply turned-on to initiate the soliton mode locking process. A theory and experimental demonstration of the existence and substantial benefits of this new operating point are presented.  

A second issue impacting microcomb integration is the realization of repetition frequencies that are both detectable and that can be readily processed by CMOS circuits. These features are essential to realize the process of comb self-referencing that underlies many comb applications\cite{diddams2010evolving}.  While detectable repetition rates are readily available in larger table top and fiber-based combs, these rates have been much more challenging to achieve in soliton microcombs owing to excessive power consumption caused by the resulting enlarged optical mode volume. To overcome this problem, ultra-high-Q silica resonators \cite{yi2015soliton,yang2018bridging} and Damascene Si$_3$N$_4$ resonators \cite{liu2018ultralow} provide effective pathways to reduce power consumption. Even there, however, it has not been possible until the present work to generate CMOS-rate solitons using integrated pumps.

\begin{figure*}
\centering
\includegraphics{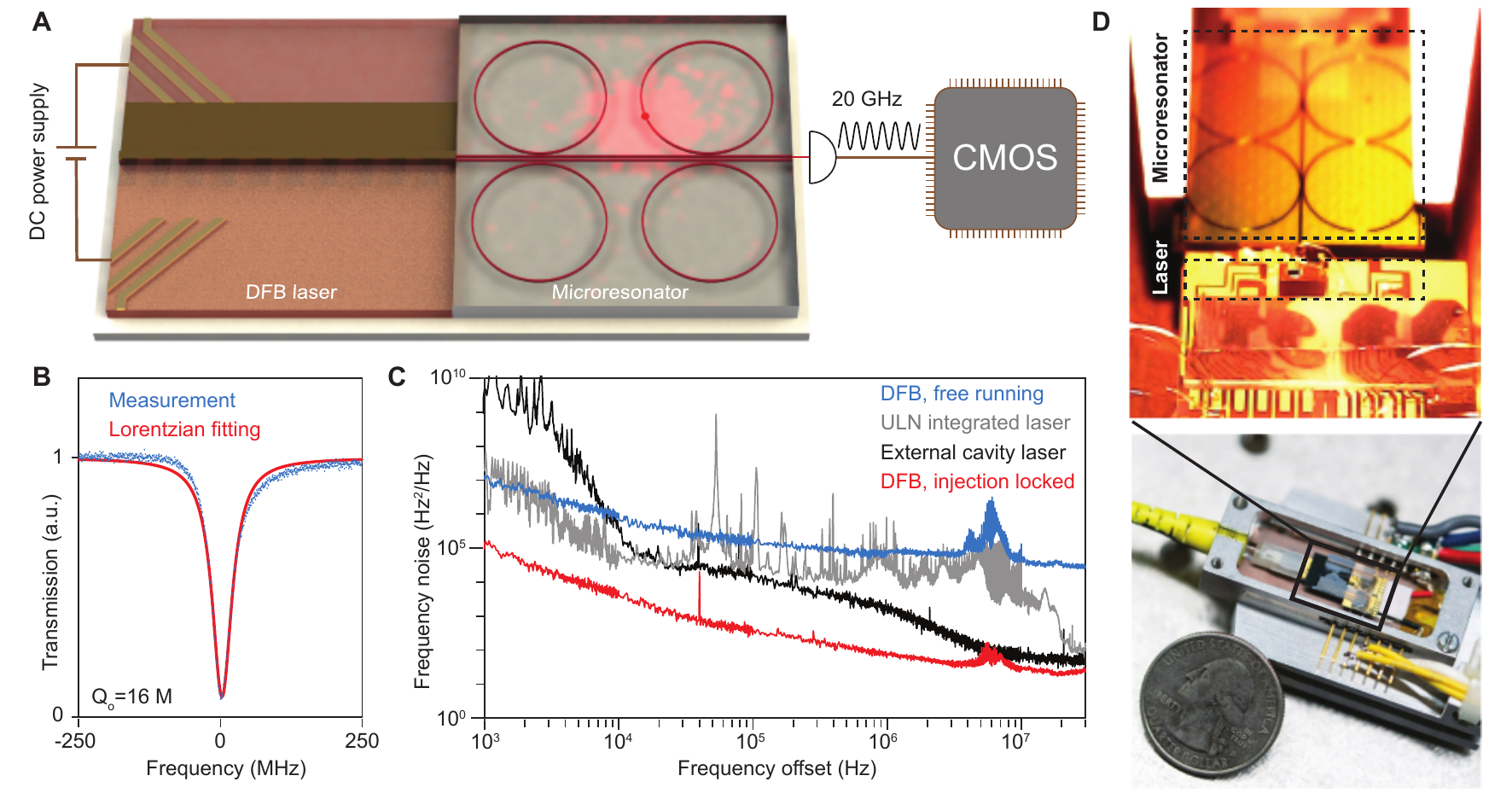}
\caption{{\bf Integrated soliton microcomb chip.} ({\bf A}) Rendering of the soliton microcomb chip that is driven by a DC power source and produces soliton pulse signals that are detectable with CMOS circuits. Four microcombs are integrated on one chip, but only one is used in these measurements. ({\bf B}) Transmission signal when scanning the laser across a cavity resonance (blue). Lorentzian fitting (red) reveals 16 million intrinsic Q factor.  ({\bf C}) Frequency noise spectral densities (SDs) of the DFB laser when it is free running (blue) and feedback-locked to a high-Q Si$_3$N$_4$ microresonator (red). For comparison, the frequency noise SDs of ultra-low-noise integrated laser on silicon (gray) and a table-top external cavity diode laser (black) are also plotted. ({\bf D}) Images of a pump/microcomb in a compact butterfly package.}
\label{fig:Fig1}
\end{figure*}

In the experiment, integrated soliton microcombs whose fabrication and repetition rate (40 GHz down to 15 GHz) are compatible with CMOS circuits \cite{abidi2014cmos} are butt-coupled to a commercial distributed-feedback (DFB) laser via inverse tapers (Fig. \ref{fig:Fig1}A).  The microresonators are fabricated using the photonic Damascence reflow process \cite{liu2018ultralow,Liu:19,supplement} and feature Q factors exceeding 16 million (Fig.\ref{fig:Fig1}B), resulting in low milliwatt-level parametric oscillation threshold, despite the larger required mode volumes of the GHz-rate microcombs. This enables chip-to-chip pumping of microcombs for the first time at these challenging repetition rates.  Up to 30 mW of optical power is launched into the microresonator. Feedback from the resonator suppresses frequency noise by around 30 dB compared with that of a free-running DFB laser (Fig. \ref{fig:Fig1}C) so that the laser noise performance surpasses state-of-the-art monolithically integrated lasers \cite{huang2019high} and table-top external-cavity-diode-lasers (ECDL). Given its compact footprint and the absence of control electronics, the pump-laser/microcomb chip set was mounted into a butterfly package (Fig. \ref{fig:Fig1}D) to facilitate measurements and also enable portability.

\begin{figure*}
\centering
\includegraphics{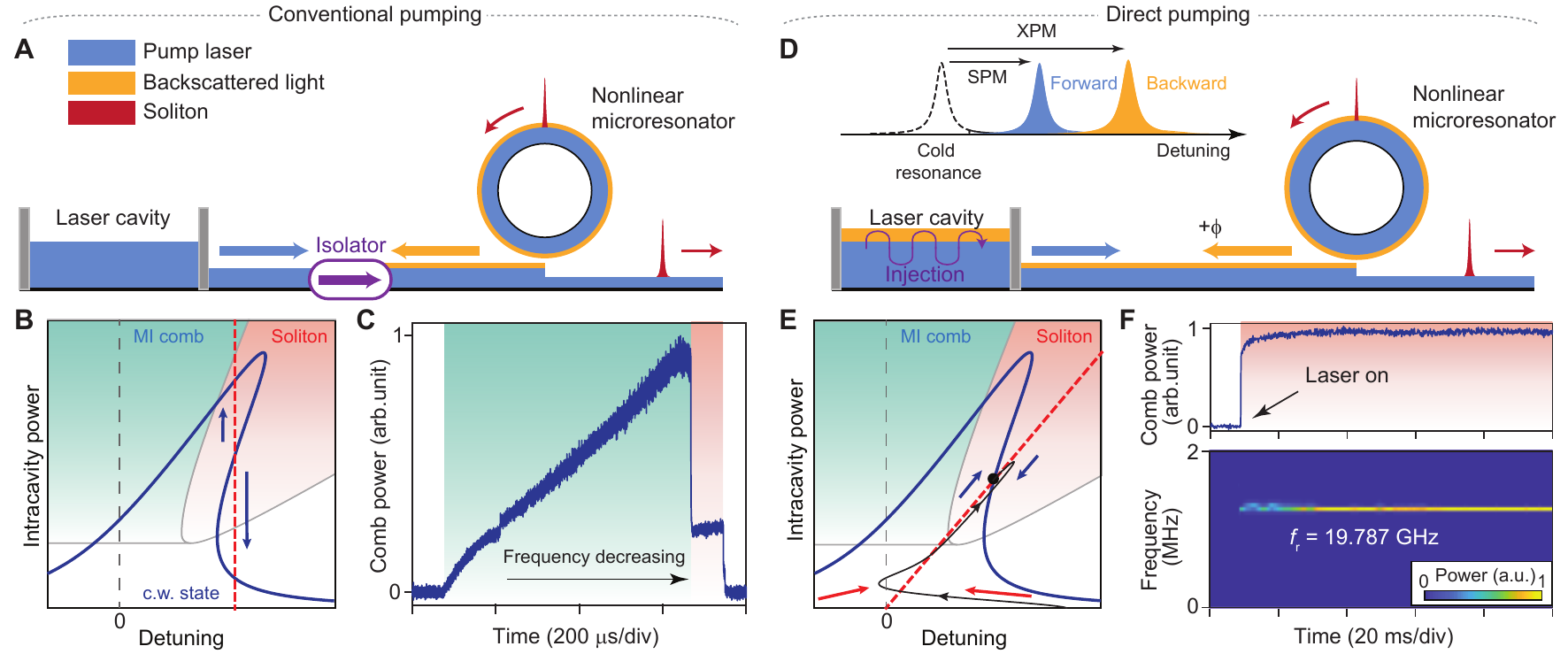}
\caption{{\bf The turnkey operating point.}
({\bf A}) Conventional soliton microcomb operation using a tunable c.w. laser. An optical isolator blocks the back-scattered light from the microresonator. ({\bf B}) Phase diagram, hysteresis curve and dynamics of the microresonator pumped as shown in panel A. The blue curve is the intracavity power as a function of cavity-pump frequency detuning.  Laser tuning (dashed red line) accesses multiple equilibria (middle point is unstable as indicated by blue arrows). ({\bf C}) Measured evolution of comb power pumped by an isolated, frequency-scanned ECDL. The step in the trace is a characteristic feature of soliton formation. ({\bf D}) Turnkey soliton microcomb generation. Non-isolated operation allows back-scattered light to be injected into the pump laser cavity . Resonances are red-shifted due to self-phase modulation (SPM) and cross-phase modulation (XPM). ({\bf E}) Phase diagram, hysteresis curve and dynamics of pump/microresonator system. A modified laser tuning curve (dashed red line) intersects the intracavity power curve (blue) to establish a new operating point from which solitons form. The feedback phase $\phi$ is set to 0 in the plot. Simulated evolution upon turning-on of the laser at a red detuning outside the soliton regime but within the locking bandwidth is plotted (solid black curve). ({\bf F}) Measured comb power (upper panel) and detected soliton repetition rate signal (lower panel) with laser turn-on indicated at 10 ms.
}
\label{fig:Fig2}
\end{figure*}

In conventional pumping of microcombs,  the laser is optically-isolated from the downstream optical path so as to prevent feedback-induced interference (Fig. \ref{fig:Fig2}A).  And on account of strong high-Q-induced resonant build-up and the Kerr nonlinearity, the intracavity power as a function of pump-cavity detuning features a bistability. The resulting dynamics can be described using a phase diagram comprising continuous-wave (c.w.), modulation instability (MI) combs and soliton regimes that are accessed as the pump frequency is tuned across a cavity resonance (Fig. \ref{fig:Fig2}B)\cite{herr2014temporal}. This tuning through the MI regime functions to seed the formation of soliton pulses. On account of the thermal hysteresis \cite{carmon2004dynamical} and the abrupt intracavity power discontinuity upon transition to the soliton regime (Fig. \ref{fig:Fig2}C), delicate tuning waveforms \cite{herr2014temporal,joshi2016thermally} or active capturing techniques \cite{yi2016active} are essential to compensate thermal transients, except in cases of materials featuring effectively negative thermo-optic response \cite{he2019self}.

Now consider removing the optical isolation as shown in Fig. \ref{fig:Fig2}D so that backscatter feedback occurs. In prior feedback locking studies the resonator has been considered to be linear so that the detuning between the feedback-locked laser and the cavity resonance is determined by the phase $\phi$ accumulated in the feedback path \cite{kondratiev2017self}. However, here the nonlinear behavior of the microresonator is included resulting in a dramatic effect on the operating point. Including the nonlinear behavior causes the resonances to be red-shifted by intensity-dependent self- and cross-phase modulation.  The relationship between detuning $\delta\omega$ (i.e. the difference of cavity resonance and laser frequency) and intracavity power of the pump mode $P_0$ can then be shown \cite{supplement} to be approximately given by,
\begin{equation}
\frac{\delta\omega}{\kappa/2}=\tan\frac{\phi}{2}+\frac{3}{2}\frac{P_0}{P_{\mathrm{th}}}
\label{Eq1}
\end{equation}
where $\kappa$ is the power decay rate of the resonance and $P_{\mathrm{th}}$ is the parametric oscillation threshold for intracavity power. This dependence of detuning on intracavity power gives rise to a single operating point at the intersection of Eq. (1) and the soliton hysteresis as shown in Fig. \ref{fig:Fig2}E. Control of the feedback phase shifts the x-intercept of Eq. (1) and thereby adjusts the operating point. Curiously, this equilibrium is stable \cite{supplement} but can lie on the middle branch of the multi-stability curve, which is unstable under conditions of conventional soliton pumping (Fig. \ref{fig:Fig2}B).

In the Supplement \cite{supplement} it is shown that the system converges to this equilibrium once the laser frequency is within a locking bandwidth (estimated to be 5 GHz in the present case). As verified both numerically (Fig. \ref{fig:Fig2}E) and experimentally (Fig. \ref{fig:Fig2}F), this behavior enables soliton mode locking by simple power-on of the pump laser (i.e., no triggering or complex tuning schemes).  Indeed, an experimental trace of the comb power shows that a steady soliton power plateau is reached immediately after turn-on of the laser. And the stable soliton emission is further confirmed by monitoring the real-time evolution of the soliton repetition rate signal (Fig. \ref{fig:Fig2}F). 
Numerical simulation of this startup process is provided in the Supplement \cite{supplement}. Finally, using this form of soliton turn-on, the thermal nonlinearity does not frustrate soliton generation and the system is highly robust with respect to temperature and environmental disturbances. Indeed, soliton generation without any external feedback control was possible for several hours in the laboratory.

\begin{figure*}[t!]
\centering
\includegraphics{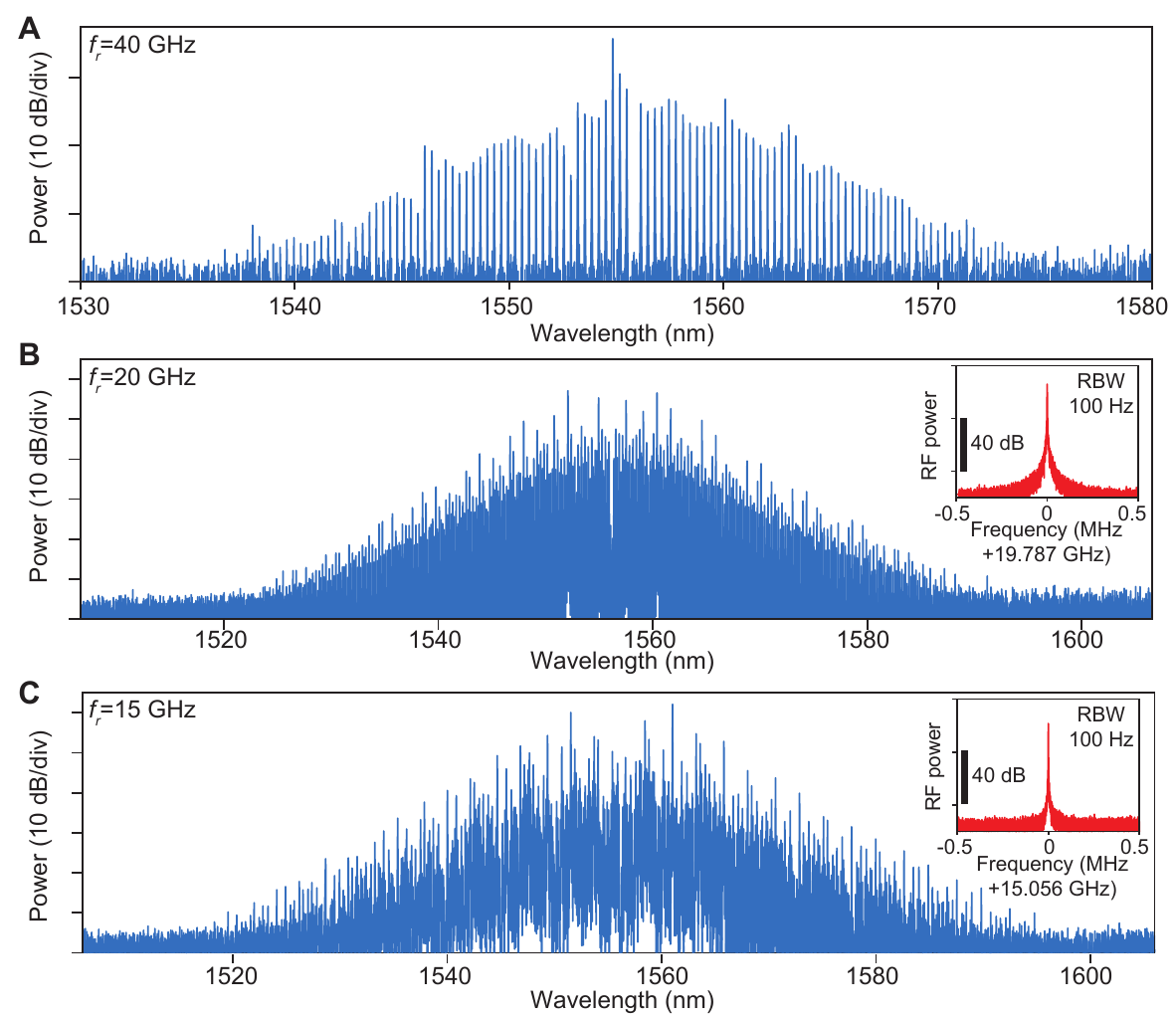}
\caption{{\bf Optical and electrical spectra of solitons.} ({\bf A}) The optical spectrum of a single soliton state with repetition rate $f_r=40$ GHz.  
({\bf B})-({\bf C}) Optical spectra of multi-soliton states at 20 GHz and 15 GHz repetition rates. Insets: Electrical beatnotes showing the repetition rates.}
\label{fig:Fig3}
\end{figure*}

Figure \ref{fig:Fig3} shows the optical spectra of a single-soliton state with 40 GHz repetition rate and multi-soliton states with 20 GHz and 15 GHz repetition rates. The pump laser at 1556 nm is attenuated at the output by a fiber-Bragg-grating notch filter in these spectra. The coherent nature of these soliton microcombs is confirmed by photodetection of the soliton pulse streams, and reveals high-contrast, single-tone electrical signals at the indicated repetition rates. Numerical simulations have confirmed the tendency of turnkey soliton states consisting of multiple solitons, which is a direct consequence of the high intracavity power and its associated MI gain dynamics \cite{supplement}. However,  single-soliton operation is accessible for a certain combination of pump power and feedback phase.

\begin{figure*}[t!]
\centering
\includegraphics{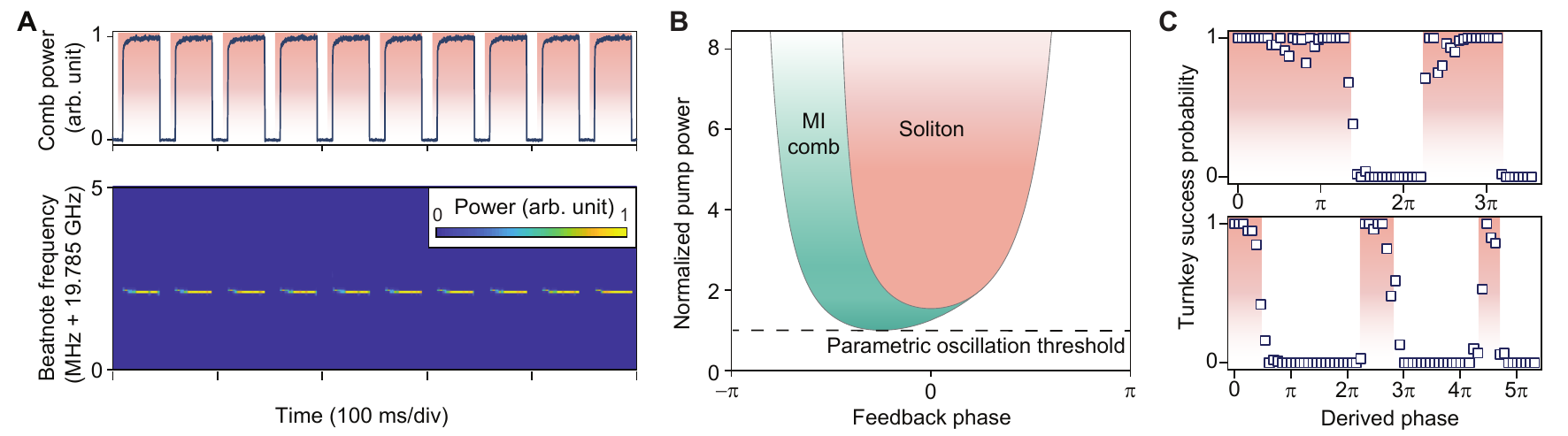}
\caption{{\bf Demonstration of turnkey soliton generation.} ({\bf A}) 10 consecutive switching-on tests are shown.  The upper panel gives the measured comb power versus time. The laser is switched on periodically as indicated by the shaded regions. The lower panel is a spectrogram of the soliton repetition rate signal measured during the switching process. ({\bf B}) Phase diagram of the integrated soliton system with respect to feedback phase and pump power. The pump power is normalized to the parametric oscillation threshold. ({\bf C}) Turnkey success probability versus relative feedback phase of 20 GHz (upper panel) and 15 GHz (lower panel) devices.  Each data point is acquired from 100 switch-on attempts.}
\label{fig:Fig4}
\end{figure*}

To demonstrate repeatable turnkey operation, the laser current is modulated to a preset current by a square wave to simulate the turn-on process. Soliton microcomb operation is reliably achieved as confirmed by monitoring soliton power and the single-tone beating signal (Fig. \ref{fig:Fig4}A). More insight into the nature of the turnkey operation is provided by the phase diagram near the equilibrium point for different feedback phase and pump power (Fig. \ref{fig:Fig4}B). The turnkey regime occurs above a threshold power within a specific range of feedback phases. Moreover, the regime recurs at 2$\pi$ increments of feedback phase, which is verified experimentally (Fig. \ref{fig:Fig4}C). Consistent with the phase diagram, a binary-like behavior of turn-on success is observed as the feedback phase is varied. In the measurement the feedback phase was adjusted by control of the gap between the facets of the laser and the microcomb bus waveguide. A narrowing of the turn-on success window with an increased feedback phase is believed to result from the reduction of the pump power in the bus waveguide with increasing tuning gap (consistent with Fig.\ref{fig:Fig4}B).

Besides the physical significance and practical impact of the new operating point, our demonstration of a turnkey operating regime is an important simplification of soliton microcomb systems.   Moreover, the application of this method in an integrated CMOS-compatible system represents a milestone towards mass production of optical frequency combs. The microwave-rate soliton microcombs generated in these butterfly packaged devices will benefit several comb applications including miniaturized frequency synthesizers \cite{spencer2018optical} and optical clocks \cite{newman2019architecture}. Moreover, the recent demonstration of low power comb formation in III-V microresonators \cite{chang2019ultra} suggests that monolithic integration of pumps and soliton microcombs is feasible using the methods developed here. A phase section could be included therein or in advanced versions of the current approach to electronically control the feedback phase.  Finally, due to its simplicity, this approach could be applied in other integrated high-Q microresonator platforms \cite{yang2018bridging,he2019self,gong2018high} to attain soliton microcombs across a wide range of wavelengths.

\medskip
\noindent {\bf Acknowledgments} \\
The authors thank Gordon Keeler, Scott Papp, Travis Briles, Justin Norman, Minh Tran for fruitful discussions, Yeyu Tong, Songtao Liu for assistance in characterizations, and Freedom Photonics for providing the lasers. 

\noindent{\bf Funding:} Supported by the Defense Advanced Research Projects Agency (DARPA) under DODOS (HR0011-15-C-055) programs, Microsystems Technology Office (MTO). 

\noindent{\bf Author contributions:} B.S., L.C., Q.-F.Y., J.L., J.B. and K.V. conceived the experiment. D.K., L.C., B.S. and Q.-F.Y. packaged the chip. J.L., R.N.W., J.H. and T.L. designed, fabricated and tested the Si$_3$N$_4$ chip devices. H.W. constructed the theoretical model. Measurements were performed by B.S., L.C., Q.-F.Y. with assistance from H.W., C.X., W.Q.X., J.G., L.W. and Q.-X.J. All authors analyzed the data and contributed to writing the manuscript. J.B., K.V. and T.J.K. supervise the project and the collaboration.

\noindent{\bf Competing interests:} The authors declare no competing financial interests.


\clearpage
\onecolumngrid
\appendix
\renewcommand{\theequation}{S\arabic{equation}}
\renewcommand{\thefigure}{S\arabic{figure}}
\setcounter{figure}{0}
\setcounter{equation}{0}

\section{Silicon nitride chip fabrication}
The Si$_3$N$_4$ \cite{Moss:13} chip devices were fabricated using the photonic Damascence reflow process \cite{Pfeiffer:18a}. Deep-UV stepper lithography is used to pattern the waveguides as well as the stress-release patterns \cite{Pfeiffer:18a} to prevent Si$_3$N$_4$ film cracks from the low-pressure chemical vapor deposition (LPCVD) process. The waveguide patterns are dry-etched into the SiO$_2$ substrate from the photoresist mask. The substrate is then annealed at 1250$^\circ$C at atmospheric pressure, such that SiO$_2$ reflow happens which reduces the waveguide sidewall roughness \cite{Pfeiffer:18}. Afterwards, LPCVD Si$_3$N$_4$ is deposited on the substrate, followed by chemical-mechanical polishing to planarize the substrate top surface. The substrate is further annealed to remove residual hydrogen contents in Si$_3$N$_4$, followed by thick SiO$_2$ top cladding deposition and SiO$_2$ annealing. 

\begin{figure}
\centering
\includegraphics{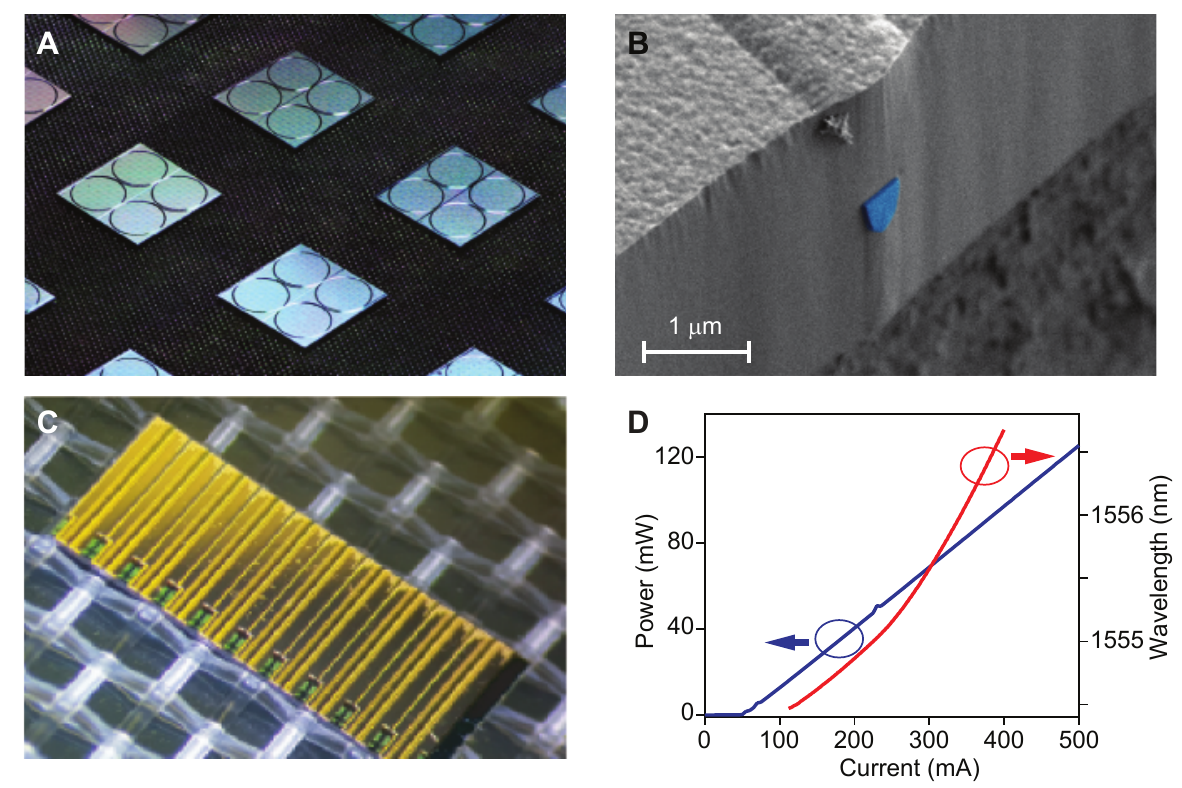}
\caption{({\bf A}) Photo of Si$_3$N$_4$ microresonator chip devices. ({\bf B}) Scanning electron microscope (SEM) image of the smooth facet of Si$_3$N$_4$ chips. The Si$_3$N$_4$ inverse taper for butt coupling is blue-colored. ({\bf C}) Microscopic image of a chip with 10 DFB lasers. ({\bf D}) Light-current curve (blue) and the wavelength response (red) of the DFB laser.}
\label{S1}
\end{figure}

To separate the wafer into chips of $5\times5$ mm$^2$ size, deep reactive ion etching (RIE) is used to define the chip facets for superior surface quality, which is particularly important in order to achieve good contact for the butt coupling between the Si$_3$N$_4$ chip and the laser chip. Commonly, dicing of SiO$_2$ and silicon together is widely used, however this method has challenges to achieve smooth SiO$_2$ facets due to the very narrow operational window of SiO$_2$ dicing. In our fabrication process, AZ 9260 photoresist of 8 $\mu$m thickness is used as the mask for the deep RIE to create chip facets. The RIE is composed of two steps: dry etch of 7-$\mu$m-thick SiO$_2$ using He/H$_2$/C$_4$F$_8$ etchants, and Bosch process to remove 200 $\mu$m silicon using SF$_6$/C$_4$F$_8$ etchants. The deep RIE thus can create a smooth chip facet for butt coupling as shown in Fig. \ref{S1}B. After the deep RIE, the wafer is diced into chips using only the silicon dicing recipe.

\section{DFB laser characterization}
The DFB pump laser fabricated by Freedom Photonics \cite{milan2014high} has high-reflection coatings at the back side facet and anti-reflection coatings at the front side facet. It has a 50 mA threshold current, and the total output power of the laser can reach 120 mW with a slope efficiency around $0.24 $ mW/mA (Fig. \ref{S1}D). The peak wall-plug efficiency is inferred to exceed 20\%. The lasing wavelength also shifts from $\sim 1554.5$ nm to $\sim 1556.5$ nm with an increase of bias current ($\sim$ -1.4 GHz/mA).

\section{Experimental details}
The laser frequency noise is obtained using the homodyne method by analyzing the electrical beat signals of two beams derived from a single laser source. One beam is sent into a long fiber delay and then combined with the other beam whose frequency has been shifted. Coherence between two beams is removed by the fiber delay. The linewidth of the laser can be inferred from the noise of the electrical beatnote signal. A fiber-Bragg-grating filter is used for picking out the single pump line when the laser is injection locked and comb is generated. The table-top ECDL used for frequency noise measurement in Fig.1B is 81608A from Keysight.

The soliton beatnote is down-mixed with a local oscillator to measure its real-time evolution. The frequency of the local oscillator is set slightly lower than the repetition rate of the solitons. A high-speed oscilloscope is used to record the trace from which the power spectrum is obtained by fast Fourier transform. The time window of the spectrograph is 20 $\mu$s and the resolution bandwidth is 50 kHz.

The relative feedback phases are estimated from the gap between the facets of the laser and the bus waveguide, which can be adjusted by a open-loop piezo micro-stepping motor (PZA12 from Newport). The step size of the actuator was calibrated by the reference mark on the chip.

\section{Theory of turnkey soliton generation}

The injection locking system consists of three parts: the soliton optical field $A_{\mathrm{S}}$, the backscattering field $A_{\mathrm{B}}$, and the laser field $A_{\mathrm{L}}$. The complete equations of motions are \cite{godey2014stability,kondratiev2017self}:

\begin{equation}
\begin{split}
\frac{\partial A_\mathrm{S}}{\partial t} &= -\frac{\kappa}{2}A_\mathrm{S}-i\delta\omega A_\mathrm{S}+i\frac{D_2}{2}\frac{\partial^2 A_\mathrm{S}}{\partial\theta^2}+i\frac{\kappa}{2} \frac{|A_\mathrm{S}|^2+2|A_\mathrm{B}|^2}{E_\mathrm{th}}A_\mathrm{S}+i\beta\frac{\kappa}{2} A_\mathrm{B}-\sqrt{\kappa_\mathrm{R}\kappa_\mathrm{L}}e^{i\phi_\mathrm{B}}A_\mathrm{L}\\
\frac{dA_\mathrm{B}}{dt} &= -\frac{\kappa}{2}A_\mathrm{B}-i\delta\omega A_\mathrm{B}+i\frac{\kappa}{2}\frac{|A_\mathrm{B}|^2+2\int_0^{2\pi}|A_\mathrm{S}|^2 d\theta/(2\pi)}{E_\mathrm{th}} A_\mathrm{B} +i\beta\frac{\kappa}{2} \overline{A_\mathrm{S}}\\
\frac{dA_\mathrm{L}}{dt} &=i\delta\omega_\mathrm{L}A_\mathrm{L}-i\delta\omega A_\mathrm{L}+\frac{g(|A_\mathrm{L}^2|)}{2}(1+i\alpha_g)A_\mathrm{L}-\sqrt{\kappa_\mathrm{R}\kappa_\mathrm{L}}e^{i\phi_\mathrm{B}}A_\mathrm{B}
\end{split}
\label{master}
\end{equation}
where the field amplitudes are normalized so that $\int_0^{2\pi}|A_\mathrm{S}|^2 d\theta/(2\pi)$, $|A_\mathrm{B}|^2$ and $|A_\mathrm{L}|^2$ are the optical energies of their respective modes, $t$ is the evolution time, $\theta$ is the resonator angular coordinate, $\kappa$ is the resonator mode loss rate (assumed to be equal for $A_\mathrm{S}$ and $A_\mathrm{B}$), $\delta\omega$ is the detuning of the cold-cavity resonance compared to injection-locked laser ($\delta\omega>0$ indicates red detuning of the pump frequency relative to the cavity frequency), $D_2$ is the second-order dispersion parameter, $E_\mathrm{th}$ is the parametric oscillation threshold for intracavity energy, $\beta$ is the dimensionless backscattering coefficient (normalized to $\kappa/2$), $\phi_\mathrm{B}$ is the propagation phase delay between the resonator and the laser, $\kappa_\mathrm{R}$ and $\kappa_\mathrm{L}$ are the external coupling rates for the resonator and laser respectively, $\omega_\mathrm{L}$ is the detuning of the cold-cavity resonance relative to the free-running laser frequency, $g$ is the net intensity-dependent gain, and $\alpha_g$ is the amplitude-phase coupling factor. The average soliton field amplitude $\overline{A_\mathrm{S}}=\int_0^{2\pi}A_\mathrm{S} d\theta/(2\pi)$ is also the amplitude on the pumped mode, and by using $\overline{A_\mathrm{S}}$ in the equation for $A_\mathrm{B}$ we have assumed that only the mode being pumped contributes significantly in the locking process, which can be justified if a single-frequency laser is used.

We will introduce some dimensionless quantities to facilitate the discussion. Define: normalized soliton field amplitude as $\psi=A_\mathrm{S}/\sqrt{E_\mathrm{th}}$, normalized amplitude on the pumped mode as $\rho=\overline{A_\mathrm{S}}/\sqrt{E_\mathrm{th}}$, normalized total intracavity power as $P=\int_0^{2\pi}|A_\mathrm{S}|^2 d\theta/(2\pi E_\mathrm{th})$, normalized detuning as $\alpha=2\delta\omega/\kappa$, normalized evolution time as $\tau=\kappa t/2$ and normalized pump as $F=-2\sqrt{\kappa_\mathrm{R}\kappa_\mathrm{L}}e^{i\phi_\mathrm{B}}A_\mathrm{L}/(\kappa\sqrt{E_\mathrm{th}})$. The equation for $A_\mathrm{S}$ can then be put into the dimensionless form
\begin{equation}
\frac{\partial \psi}{\partial \tau} = -(1+i\alpha)\psi+i\frac{D_2}{\kappa}\frac{\partial^2 \psi}{\partial\theta^2}+i \left(|\psi|^2+\frac{2|A_\mathrm{B}|^2}{E_\mathrm{th}}\right)\psi+i\beta \frac{A_\mathrm{B}}{\sqrt{E_\mathrm{th}}}+F
\end{equation}

To simplify the equations we also consider the following approximations. We assume that backscattering is weak ($\beta\ll 1$) so that nonlinearities caused by $|A_\mathrm{B}|^2$ are negligible. Also, the coupled amplitude from $A_\mathrm{B}$ to $A_\mathrm{S}$ is small and can be neglected. The propagation phase $\phi_\mathrm{B}$ depends on the precise frequency of the locked laser and material dispersion, but we assume that the feedback length is short ($L\ll c/(n\kappa)$, where $c$ is the speed of light in vacuum and $n$ is the refractive index of the material) so that $\phi_\mathrm{B}$ can be treated as constant. Finally, we assume that gain saturation is strong enough ($\partial g/\partial |A_\mathrm{L}^2|\gg\beta\gamma/|A_\mathrm{L}^2|$, where $\gamma$ is the laser mode loss rate) so that the laser operates at constant power ($|A_\mathrm{L}|$ is constant) and is not affected by the feedback.

With these approximations, the stationary solution for $A_\mathrm{B}$ can be found as
\begin{equation}
A_\mathrm{B}=\frac{i\beta\rho}{1+i(\alpha-2P)}\sqrt{E_\mathrm{th}}
\end{equation}
and the stationary equation for $A_\mathrm{L}$ reduces to an algebraic equation for $\delta\omega$ (i.e., $\alpha=2\delta\omega/\kappa$) after eliminating the gain $g$,
\begin{equation}
\alpha = \alpha_\mathrm{L}-\frac{2}{\kappa}\mathrm{Im}\left[\sqrt{\kappa_\mathrm{R}\kappa_\mathrm{L}}e^{i\phi_\mathrm{B}}(1-i\alpha_g)\frac{A_\mathrm{B}}{A_\mathrm{L}}\right] = \alpha_\mathrm{L}+\frac{4\kappa_\mathrm{R}\kappa_\mathrm{L}}{\kappa^2}\mathrm{Im}\left[e^{2i\phi_\mathrm{B}}\frac{i\beta}{1+i(\alpha-2P)}(1-i\alpha_g)\frac{\rho}{F}\right]
\end{equation}
where $\mathrm{Im}[\cdot]$ is the imaginary part function and we have defined the normalized detuning for the free-running laser $\alpha_\mathrm{L}=2\delta\omega_\mathrm{L}/\kappa$.

Based on these equations, We define the locking strength, which is proportional to the locking bandwidth,
\begin{equation}
K=\frac{4\kappa_\mathrm{R}\kappa_\mathrm{L}}{\kappa^2}|\beta|\sqrt{1+\alpha_g^2}
\end{equation}
and the feedback phase 
\begin{equation}
\phi=2\phi_\mathrm{B}-\mathrm{arctan}(\alpha_g)+\mathrm{Arg}[\beta]+\frac{\pi}{2}
\end{equation}
where $\mathrm{Arg}[\cdot]$ is the argument function and the $\pi/2$ is added for later convenience. Now the above equations can be reduced to a conventional Lugiato-Lefever equation with an additional algebraic equation that describes the dynamics of injection locking: 
\begin{equation}
\begin{split}
\frac{\partial\psi}{\partial \tau} &= -(1+i\alpha)\psi+iD_2\frac{\partial^2 \psi}{\partial\theta^2}+i |\psi|^2\psi+F\\
\alpha &= \alpha_\mathrm{L}+K\mathrm{Im}\left[e^{i\phi}\frac{1}{1+i(\alpha-2P)}\frac{\rho}{F}\right]
\end{split}
\end{equation}

It is known that combs and solitons will emerge from a continuous-wave (CW) background when its power exceeds the parametric oscillation threshold ($|\rho|>1$), and it is desirable to first study the CW excitation of the system by setting $D_2=0$. In this case, the Lugiato-Lefever partial differential equation reduces to an ordinary differential equation with $\psi=\rho$ and $P=|\rho|^2$. The steady state solution can be found from
\begin{equation}
\label{LLE_CW_sol}
F=[1+i(\alpha-|\rho|^2)]\rho
\end{equation}
and the locking equilibrium reduces to
\begin{equation}
\alpha = \alpha_\mathrm{L}+K\mathrm{Im}\left[e^{i\phi}\frac{1}{1+i(\alpha-2P)}\frac{1}{1+i(\alpha-P)}\right]=\alpha_\mathrm{L}+K\chi(P,\alpha,\phi)
\end{equation}
where we have defined the CW locking response function:
\begin{equation}
\chi(P,\alpha,\phi)=\frac{(3P-2\alpha)\cos\phi+(1-2P^2+3P\alpha-\alpha^2)\sin\phi}{[1+(\alpha-P)^2][1+(\alpha-2P)^2]}
\end{equation}

To obtain analytical results, we will also make the approximation of infinite locking bandwidth limit (i.e. $K\rightarrow\infty$). The locking condition is then equivalent to setting the locking response function to zero:
\begin{equation}
\label{lock_resp_sol}
(3P-2\alpha)\cos\phi+(1-2P^2+3P\alpha-\alpha^2)\sin\phi=0
\end{equation}

\begin{figure}
\centering
\includegraphics{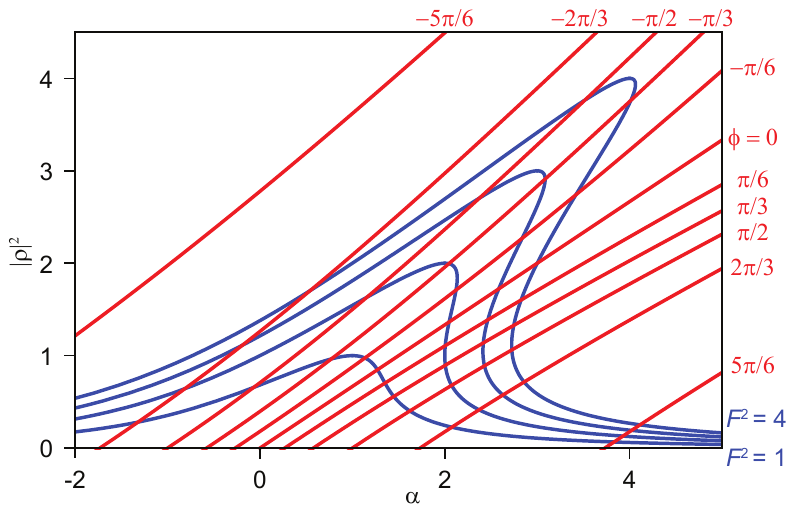}
\caption{\textbf{Continuous-wave states of the injection-locked nonlinear resonator.} Resonator characteristics are shown as the blue curves, with $F^2=1$ (bottom) to $4$ (top). Laser locking characteristics are shown as the red curves, with $\phi=-5\pi/6$ (upper left) to $5\pi/6$ (lower right).}
\label{S2}
\end{figure}

Fig. \ref{S2} shows a plot for Eq. (\ref{LLE_CW_sol}) with different pumping powers $F$ and Eq. (\ref{lock_resp_sol}) with different feedback phases $\phi$. The intersecting point of the two curves gives the CW steady state of the cavity. Note that there are two solutions to the quadratic equation Eq. (\ref{lock_resp_sol}). Only the solution branch that satisfies $\partial\chi/\partial\alpha<0$ is plotted; these are the stable locking solutions. The opposite case $\partial\chi/\partial\alpha>0$ drives the frequency away from the equilibrium.

When a resonator is pumped conventionally, the intracavity power $P$ will approach its equilibrium given by Eq. (\ref{LLE_CW_sol}). In the case of feedback-locked pumping, such power changes will also have an effect on the locking response function $\chi$, pulling the detuning to the new locking equilibrium as well (Fig. 2B and 2E in main text). A special case is $\phi=0$, where the locking equilibrium can be simply described as
\begin{equation}
\alpha=\frac{3}{2}P
\end{equation}
i.e. the detuning is pulled away from the cold cavity resonance, and the effect is exactly $3/2$ times what is expected from the self phase modulation. This is an average effect of the self phase modulation on the soliton mode and the cross phase modulation of the backscattered mode from the soliton mode. More generally, the detuning can be solved in terms of $P$ as
\begin{equation}
\alpha=\frac{3}{2}P-\cot\phi+\frac{\sqrt{4+P^2\sin^2\phi}}{2\sin\phi}
\end{equation}
where again only the solution satisfying $\partial\chi/\partial\alpha<0$ is given. Neglecting the higher-order $P^2\sin^2\phi$ term inside the square root results in a lowest order approximation:
\begin{equation}
\alpha=\mathrm{tan}\frac{\phi}{2}+\frac{3}{2}P
\end{equation}
which splits into two additive contributions: one from the feedback phase and the other from the averaged nonlinear shift. This is Eq. 1 in the main text when written using dimensional quantities.

When the dispersion term is considered, the CW solution is no longer stable, which leads to the formation of modulational instability (MI) combs. These combs will evolve into solitons if the CW state is also inside the multistability region of the resonator dynamics. By adjusting the pump power $F^2$ and feedback phase $\phi$, we can change the operating point of the cavity, and map the possible comb states to a phase diagram with $F^2$ and $\phi$ as parameters (Fig. 4B in the main text). It should be noted that the formation of combs will change the intracavity power ($P>|\rho|^2$) as well as the response function, which will slightly shift the operating point of the system. Contrary to conventional pumping phase diagrams (with $F^2$ and $\alpha$ as parameters), where soliton existence regions only implies the possible formations of solitons due to multistability, the soliton existence region here guarantees the generation of solitons as the system bypasses the chaotic comb region before the onset of MI.

\begin{figure}
\centering
\includegraphics{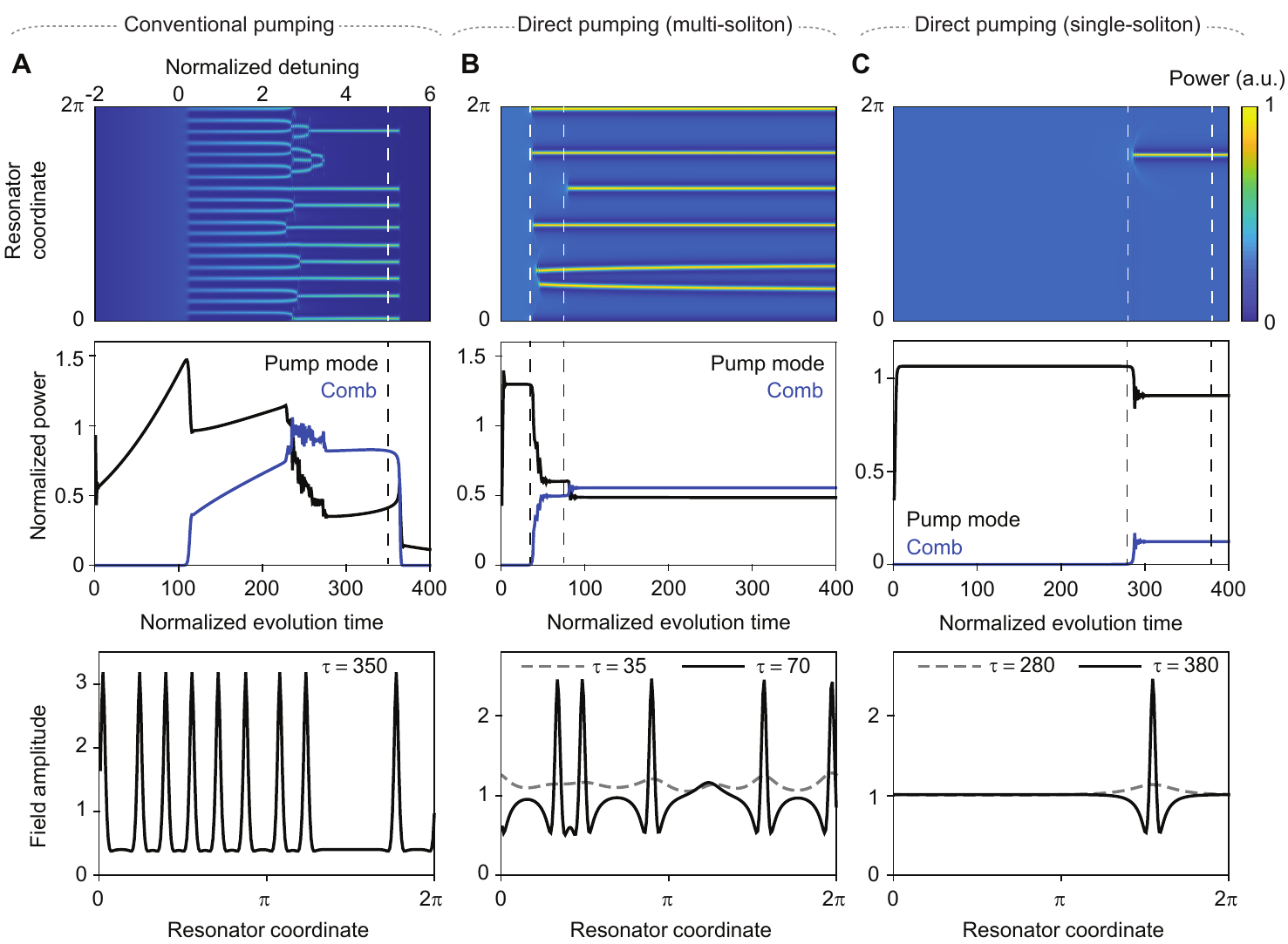}
\caption{\textbf{Numerical simulations of turnkey soliton generation. }({\bf A}) Conventional solitons are generated by sweeping the laser frequency. Parameters are $K=0$ (no feedback) and $F^2=4$. The normalized laser frequency is swept from $\alpha_\mathrm{L}=-2$ to $\alpha_\mathrm{L}=6$ within a normalized time interval of $400$. Upper panel: Soliton field power distribution as a function of evolution time and coordinates. Middle panel: Dynamics of the pump mode power (black) and comb power (blue). Lower panel: A snapshot of the soliton field at evolution time $\tau=350$ ($\alpha_\mathrm{L}=5$), also marked as a white dashed line in the upper panel and a black dashed line in the middle panel. ({\bf B}) Multiple solitons are generated via feedback locking. Parameters are $K=15$, $\phi=0.15\pi$, $F^2=3$ and $\alpha_\mathrm{L}=4.8$. Upper and middle panels are the same as in (A). Lower panel: Snapshots of the soliton field at evolution time $\tau=35$ (gray dashed line) and $\tau=70$ (black solid line), also marked as white dashed lines in the upper panel and black dashed lines in the middle panel. ({\bf C}) A single soliton is generated via injection locking. Parameters are $K=15$, $\phi=0.3\pi$, $F^2=3$ and $\alpha_\mathrm{L}=4.8$. Upper and middle panels are the same as in (A). Lower panel: Snapshots of the soliton field at evolution time $\tau=280$ (gray dashed line) and $\tau=380$ (black solid line), also marked as white dashed lines in the upper panel and black dashed lines in the middle panel.}
\label{S3}
\end{figure}

We have also performed numerical simulations to verify the above analyses (Fig. \ref{S3}). The simulation numerically integrates Eq. \ref{master} with a split-step Fourier method while fixing the amplitude of $|A_\mathrm{L}|$. Noise equivalent to about one-half photon per mode is injected into $A_\mathrm{S}$ to provide seeding for comb generation. Parameters common to all simulation cases are $D_2/\kappa=0.015$ and $|\beta|=0.05$, while others are varied across different cases and can be found in the caption of \ref{S3}. In the first case (Fig. \ref{S3}A), conventional soliton generation by sweeping the laser frequency is presented, showing the dynamics of a random noisy comb waveform collapsing into soliton pulses. This is in contrast to the turnkey soliton generation in the second case (Fig. \ref{S3}B), where solitons directly ``grow up'' from ripples in the background. Such ripples are generated by MI in those sections of the resonator with local intracavity power above the threshold. Each peak in the ripples corresponds to one soliton if collisions and other events are not considered. The process of growing solitons out of the background will continue until there is no space for new solitons or when the background falls below the MI threshold, and such dynamics explain the tendency of the turnkey soliton state to consist of multiple solitons. By carefully tuning the phase and controlling the MI gain, it is still possible to obtain a turnkey single soliton state, as shown in the third case (Fig. \ref{S3}C).

\section{Additional measurements}
\subsection{Different types of microcombs in the injection locking system}
There are several interesting solutions other than stable solitons that can be derived from the regular Lugiato-Lefever equation (LLE) \cite{godey2014stability}. One is the breather soliton, which is the type of soliton whose shape oscillates in time. Another example is the chaotic comb, which corresponds to the unstable Turing patterns or soliton state as the pump power is increased. In addition, solitons can be self-organized and form an equidistant pulse train in the microresonator, which is called a soliton crystal.

\begin{figure}
\centering
\includegraphics{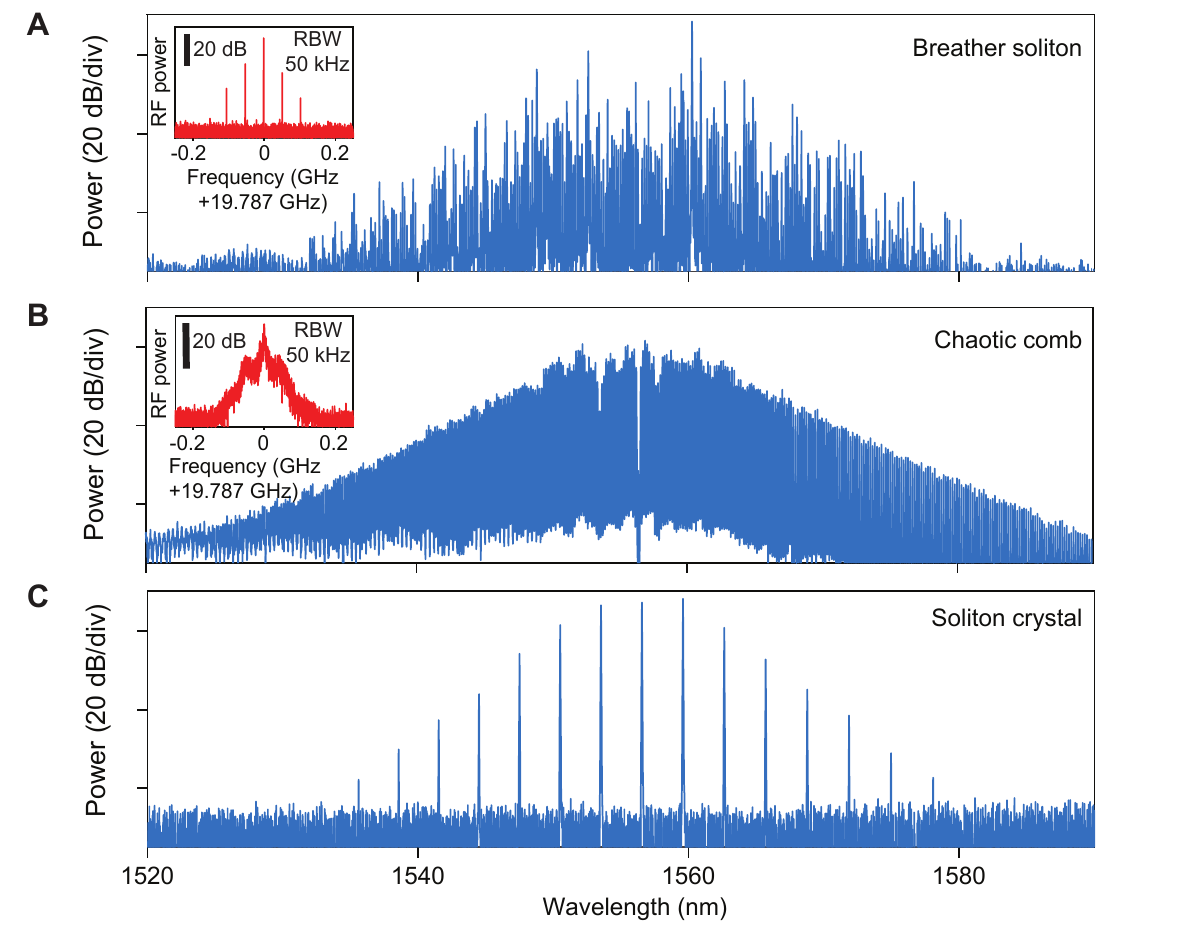}
\caption{\textbf{Optical and electrical spectra of different microcomb types.} ({\bf A})-({\bf B}) Optical spectra of breather solitons and a chaotic comb. Inset: Electrical beatnote signals. ({\bf C}) Optical spectrum of a soliton crystal state.}
\label{S4}
\end{figure}

\begin{figure}
\centering
\includegraphics{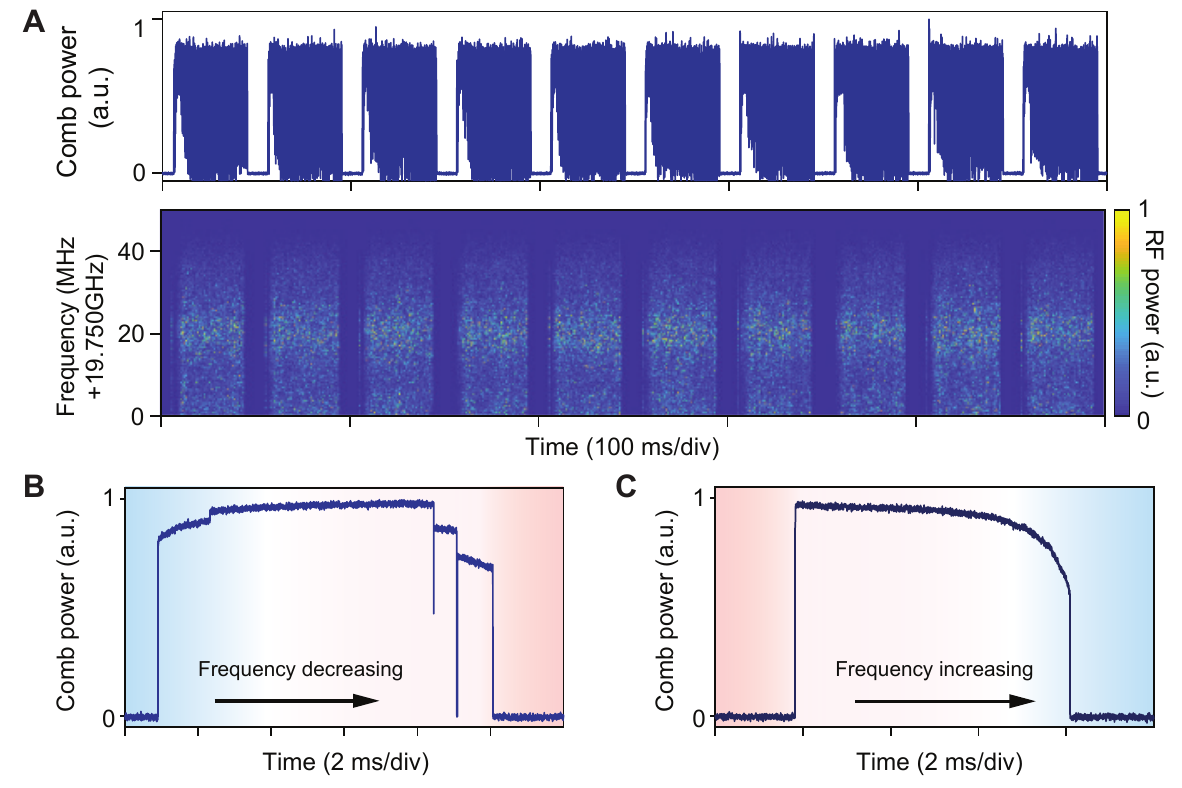}
\caption{({\bf A}) Turnkey genertion of a chaotic comb. Upper panel: Comb power evolution. Lower panel: Spectrograph of RF beatnote power. ({\bf B})-({\bf C}) Comb power evolution when the pump laser frequency is driven from blue to red ({\bf B}) and red to blue ({\bf C}) .}
\label{S5}
\end{figure}

It is also possible to operate our system in different types of microcomb states under certain feedback phase and laser driving frequency. Fig. \ref{S4} shows the experimental spectra of the breather solitons, a chaotic comb and a soliton crystal state, respectively. The turnkey generation of the chaotic comb is further shown in Fig. \ref{S5}A. The broad and noisy RF spectrum indicates that it is not mode-locked.

\subsection{Tuning of turnkey soliton microcomb system}
To further explore the performance of the turnkey soliton microcomb system, the frequency of the pump laser is driven by a linear current scan (Fig. \ref{S5}B,C). The scan speed of the driving frequency is around 0.36 GHz/ms, estimated from the wavelength-current response when the laser is free running. When the laser is scanned across the resonance, feedback locking occurs and pulls the pump laser frequency towards the resonance, until the driving frequency is out of the locking band. As shown in Fig. \ref{S5}B, the power steps indicate that soliton states with different soliton numbers can be accessed as we tune the driving current. It is worth noting that the soliton microcombs can be powered-on when the laser is scanned from red-detuned side (Fig. \ref{S5}C), which seldom happens under conventional pumping, except in cases of an effectively negative thermo-optic response system \cite{he2019self}. The comb evolution during laser scanning is a useful tool to assess the robustness of the turnkey soliton generation.

\end{document}